\documentclass[pdflatex,sn-mathphys-num]{sn-jnl}

\usepackage{graphicx}%
\usepackage{multirow}%
\usepackage{amsmath,amssymb,amsfonts}%
\usepackage{amsthm}%
\usepackage{mathrsfs}%
\usepackage[title]{appendix}%
\usepackage{xcolor}%
\usepackage{textcomp}%
\usepackage{manyfoot}%
\usepackage{booktabs}%
\usepackage{algorithm}%
\usepackage{algorithmicx}%
\usepackage{algpseudocode}%
\usepackage{listings}%

\theoremstyle{thmstyleone}%
\theoremstyle{thmstyletwo}%

\theoremstyle{thmstylethree}%

\begin{document}

\title[Multi-phase control mechanism on fibroblast dynamics]{Effects of multi-phase control mechanism on fibroblast dynamics: A segmented mathematical modeling approach}


\author[1]{\fnm{} \sur{Shuqi Fan}}\email{fanshuqi@tiangong.edu.cn}

\author[2]{\fnm{} \sur{Yuhong Zhang}}\email{yuhongzhang@tiangong.edu.cn}

\author*[1,3]{\fnm{} \sur{Jinzhi Lei}}\email{jzlei@tiangong.edu.cn}

\affil[1]{\orgdiv{School of Mathematical Sciences}, \orgname{Tiangong University}, \orgaddress{\city{Tianjin}, \postcode{300387}, \country{China}}}

\affil[2]{\orgdiv{School of Software}, \orgname{Tiangong University}, \orgaddress{\city{Tianjin}, \postcode{300387},\country{China}}}

\affil[3]{\orgdiv{Center for Applied Mathematics}, \orgname{Tiangong University}, \orgaddress{ \city{Tianjin}, \postcode{300387}, \country{China}}}


\abstract{Cell size is a fundamental determinant of cellular physiology, influencing processes such as growth, division, and function. In this study, we develop a segmented mathematical framework to investigate how different control mechanisms operating across multiple phases of the cell cycle affect fibroblast population dynamics. Building on our previous work modeling sizer, timer, and adder strategies, we extend the analysis by introducing phase-specific control schemes in the S and G2 phases, incorporating nonlinear growth dynamics and cell death. Using agent-based stochastic simulations, we examine how these mechanisms shape steady-state size distributions and respond to parameter variations. Our results reveal that the steady-state cell size distribution is primarily governed by division kernels and phase-specific control strategies, and appears remarkably robust to cell death modalities. We identify a fundamental trade-off between extrinsic and intrinsic growth feedbacks: while population-density-dependent regulation tightly limits total cell numbers, cell-size-dependent regulation acts as a proportional homeostatic mechanism, suppressing relative size variability. Furthermore, we demonstrate that population recovery is accelerated by the retention of proliferation-competent large cells. This study provides biologically relevant insights into the complex interplay between growth, division, and homeostasis, with implications for understanding tissue repair and disease progression.}





\maketitle

\section{Introduction}
\label{sec1}
Fibroblasts are key cellular constituents of mammalian connective tissues, playing essential roles in tissue development, wound healing, and homeostasis maintenance \cite{2021Fibroblast, 2023Fibroblast, franklin2021fibroblasts}. They synthesize structural components of the extracellular matrix, including collagen and elastin, and mediate physiological responses such as immune cell recruitment and angiogenesis through paracrine signaling \cite{PLIKUS20213852, davidson2021fibroblasts, lynch2018fibroblast}. While much attention has been paid to fibroblast molecular signaling, increasing evidence suggests that physical characteristics---particularly cell size---are also critical to their biological function \cite{dubois2012roles}. 

Cell size is not a fixed property but a dynamic variable integrating signals from cellular metabolism, mechanical stimuli, and ionic balance \cite{xie2024unveiling, jiang2013cellular, leal2016dynamic, mongin2001mechanisms, lin2022connecting, lee2007mtor, lengefeld2023cell, hernandez2007general}. For fibroblasts, variations in cell size correlate strongly with phenotype and function. Under homeostatic conditions, fibroblasts maintain a moderate size that supports matrix remodeling and tissue integrity \cite{mcmanus1995regulation}. During wound healing, size reduction enhances migratory ability, while subsequent size enlargement supports collagen synthesis via endoplasmic reticulum expansion. In fibrotic conditions, disruptions in size homeostasis can induce fibroblast-to-myofibroblast differentiation, contributing to pathological remodelling \cite{xu2025piezo1, 2023Serine, razdan2018telomere}. Moreover, recent studies have shown that fibroblast size heterogeneity varies by tissue origin and correlates with disease progression, reinforcing the view of size as a critical biomarker of fibroblast state \cite{2022Sphingolipids, gao2024cross, ghonim2023pulmonary,2024Cross, davidson2021fibroblasts}.

Despite its biological significance, the principles governing cell size regulation remain poorly understood. In unicellular organisms, three canonical control mechanisms---sizer, timer, and adder---have been proposed to explain size homeostasis. In yeast, the sizer model ensures division at a critical size threshold \cite{Turner:2012aa,1977Control}, while in Bacillus crescentus, the timer model sets a fixed growth duration dependent on initial size \cite{banerjee2017biphasic}. Escherichia coli follows the adder model, maintaining a constant size increment between birth and division \cite{tanouchi2015noity}. 

Interestingly, mammalian cells appear to adopt size control strategies that resemble those found in unicellular organisms. However, within the complex microenvironments of multicellular systems, these mechanisms likely operate through more intricate and synergistic regulatory networks. Cadart et al. used microtubule culture, fluorescence exclusion measurements, and cell-cycle labeling to track single-cell size trajectories, and through mathematical modeling, they demonstrated that most mammalian cells---including cancerous, non-cancerous, and progenitor lines---exhibit behavior close to the adder model \cite{Cadart:2018aa}. Similarly, Miotto et al. combined live-cell fluorescence labeling, flow cytometry, and an analytical model to show that a sizer-like strategy governs division timing in T cells, with division intervals following an Erlang distribution \cite{Miotto2024A}. 

The problem of cell size control poses a significant mathematical challenge for quantitatively modeling and analyzing the evolution of size distribution within cell populations. Over the years, mathematical frameworks based on partial differential equations (PDEs) and stochastic processes have been developed to describe the effects of various size control mechanisms \cite{Bell1968Cell,Diekmann1984On,jia2022characterizing,xia2020PDE}. These models typically describe the evolution of cell number density as a function of cell age and size, often assuming symmetric division. In our previous work \cite{Fan:2025dg}, we established a first-order PDE framework to model population-level size dynamics under the size, timer, and adder mechanisms. That study focused on global control strategies, providing both exact analytical solutions and validation through agent-based stochastic simulations. However, biological evidence suggests that size regulation is often phase-dependent, with different control mechanisms operating at distinct stages of the cell cycle. For example, while G1 duration is inversely correlated with birth size \cite{XIE2020916, dolznig2004evidence}, this alone cannot account for the long-term size maintenance of size homeostasis, implying that later phases such as S and G2 also contribute to size regulation \cite{Liu:2024aa}. 

In the present study, we focus on fibroblasts and extend our modeling approach to incorporate phase-specific control mechanisms across multiple stages of the cell cycle. We construct a segmented PDE model in which distinct regulatory strategies govern the S and G2 phases, while experimentally validated adder/timer behaviors are assumed for the G1 phase. Based on experimental data from fibroblasts, we calibrate model parameters, explicitly compare the effects of linear vs. nonlinear growth assumptions and cell death, and perform extensive agent-based stochastic simulations to explore the resulting size distributions. Our results reveal two key findings: (1) the breadth of the steady-state size distribution is primarily governed by the growth rate and division asymmetry; and (2) the specific form of the growth function (linear vs. nonlinear) is the dominant factor regulating total population size. Our goal is to elucidate how multi-phase control contributes to population-level size homeostasis and to provide mechanistic insights into the dynamic regulation of cell size in mammalian systems.

\section{Materials and methods}
\label{sec2}
\begin{figure}[htbp]
\centering
\includegraphics[width=11cm]{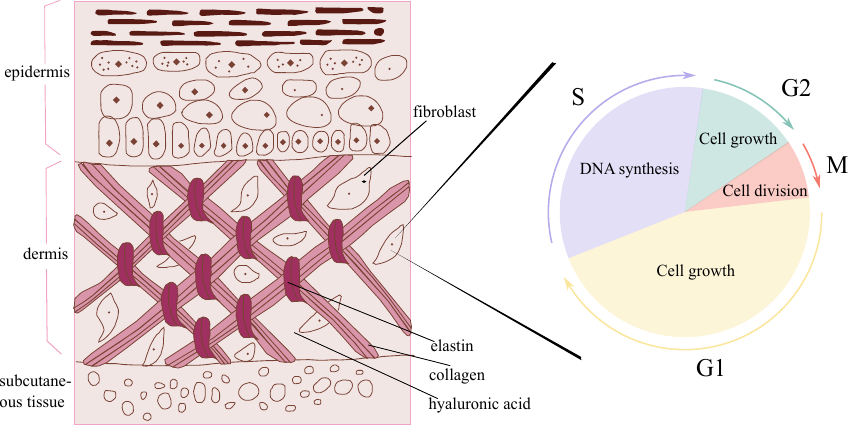}
\caption{Fibroblasts in the skin. The skin consists of the epidermis, dermis, and subcutaneous tissue. Fibroblasts are primarily located in the extracellular matrix of the dermis, where they exhibit stellate or spindle-shaped morphologies with multiple cytoplasmic protrusions that interact with surrounding collagen, elastin, and neighboring cells (e.g., immune cells and vascular endothelial cells). The right panel shows a schematic of the fibroblast cell cycle.}
\label{fig:1}
\end{figure}

In this study, we develop a phase-segmented mathematical model to characterize the growth, division, and death dynamics of mammalian fibroblasts and to investigate the evolution of cell size distributions at the population level (Fig. \ref{fig:1}). Fibroblast behavior is tightly regulated by the cell cycle, which orchestrates cell growth, division, and death in a highly coordinated manner. These processes are further influenced by the surrounding microenvironment. 

During the growth phase, fibroblasts actively synthesize macromolecules and increase in size through the accumulation of cellular components. Mitosis involves a precisely regulated sequence of events in which a mother cell divides to produce two daughter cells, each inheriting identical nuclear material and comparable cytoplasmic content. In addition to proliferation, cell death is also phase-dependent: the probability of apoptosis varies across different stages of the cell cycle, with cells exhibiting distinct sensitivities depending on their phase.

To capture these biological features, we propose a mathematical framework that segments the cell cycle into distinct regulatory phases, allowing for differential modeling of growth, division, and death. The details of the modeling approach are described below. 

\subsection{Cell cycle segmentation model}
\begin{figure}[htbp]
\centering
\includegraphics[width=11cm]{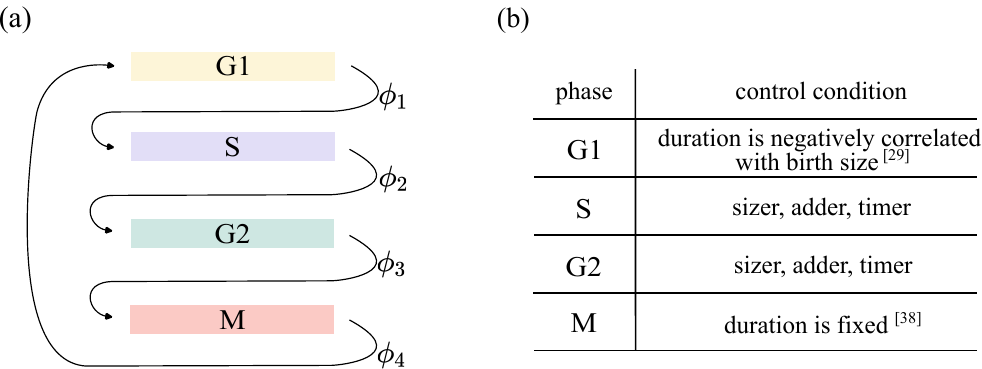}
\caption{Control mechanisms across cell cycle phases. (a) Phase progression regulated by control conditions $\phi_i$, where $i = 1, 2, 3, 4$ correspond to the G1, S, G2, and M phases, respectively. A cell advances to the next phase only when the corresponding $\phi_i$ condition is satisfied. (b) Specific regulatory mechanisms for each phase.}
\label{fig:2}
\end{figure}

We consider a population of mammalian fibroblasts and construct a phase-segmented model to capture the regulatory mechanisms operating during the G1, S, G2, and M phases of the cell cycle (Fig. \ref{fig:2}a). Each phase is associated with a distinct control condition, denoted by $\phi_i$ for $i = 1, 2, 3, 4$, corresponding to the G1, S, G2, and M phases, respectively. A cell can progress from the $i$-th phase to the next only when the control condition $\phi_i$ is satisfied. 

Based on the above assumptions, the dynamics of the cell population are governed by the following system of partial differential equations:
\begin{eqnarray}
\label{eq:1}
\dfrac{\partial n_1 (s, a, t)}{\partial t}  &=& - \nabla_1 n_1(s, a, t) - (\mu_1(s, a) + \phi_1 (s, a))n_1(s, a, t),\\
\label{eq:2}
\dfrac{\partial n_i (s, a, t)}{\partial t}  &=& - \nabla_i n_i(s, a, t) - (\mu_i(s, a) + \phi_i (s, a))n_{i}(s, a, t) \nonumber\\
&&{}+ \phi_{i-1}(s, a)n_{i-1}(s, a, t),\qquad i = 2, 3, 4,
\end{eqnarray}
where $\nabla_i$ represents an age-structured operator corresponding to a growth rate $v_i(s, a)$:
\begin{equation}
\label{eq:nai}
\nabla_i n_i(s, a, t) = \dfrac{\partial n_{i}(s, a, t)}{\partial a_i} + \dfrac{\partial (v_{i} (s, a)n_{i}(s, a, t))}{\partial s}. 
\end{equation}
The detailed derivation is provided in Appendix \ref{sec:Derivation_eq}. The boundary condition is given by (see  Appendix \ref{sec:Derivation_BC}):
\begin{equation}
\label{eq:3}
n_1(s, 0, t) = 2 \displaystyle\int_0^{+\infty}\int_0^{+\infty} p(s, s')\phi_4(s', a)n_4(s', a, t)\mathrm{d}a \mathrm{d}s'.
\end{equation}
Here, $n_i(s, a, t)$ denotes the number of cells in phase $i$ with size $s$ and age $a$ at time $t$. In our framework, the age variable $a$ represents the time elapsed since the cell's birth (entry into G1) and is effectively inherited as the cell transitions between phases. The functions $v_i(s, a)$, $\phi_i(s, a)$, and $\mu_i(s, a)$ represent the cell growth rate, phase-specific control function, and death rate, respectively. 

The function $p(s, s')$ denotes the probability density that a mother cell of size $s'$ divides into a daughter cell of size $s$, satisfying the normalization condition:
$$\int_0^{+\infty} p(s, s') \mathrm{d}s = 1,\quad \forall s'.$$
The factor $2$ in Eq. \eqref{eq:3} accounts for the production of two daughter cells per division event. Biologically, cell division typically involves the partitioning of cytoplasmic volume, where daughter cells inherit a proportion of the mother cell's size rather than a fixed absolute volume. Empirical studies suggest that this size allocation ratio typically follows a Beta distribution \cite{KoppesGrover:1992relationship}. To simplify our analysis, we assume that the division kernel depends only on the size ratio $u = s/s'$, such that $p(s, s') = \frac{1}{s'}p(s/s')$. Specifically, we define $p(u)$ as:
\begin{equation}
p(u) = \dfrac{u^{\alpha-1}(1-u)^{\beta-1}}{B(\alpha,\beta)},\quad B(\alpha,\beta) = \dfrac{\Gamma(\alpha)\Gamma(\beta)}{\Gamma(\alpha + \beta)},
\end{equation}
where $\Gamma(\cdot)$ is the gamma function. 

Let $f_i(s, a, t)$ denote the normalized cell size distribution in phase $i$, defined by
\begin{equation}
f_{i} (s, a, t)= \dfrac{n_i(s, a, t)}{N(t)},\  N(t) = \sum_{i = 1}^4 N_i(t), \ N_i(t) =  \int_0^{+\infty}\int_0^{+\infty} n_i(s, a, t) \mathrm{d}a\mathrm{d} s.  
\end{equation}
Then the governing equations for $f_i$ are:
\begin{eqnarray}
\dfrac{\partial f_1 (s, a, t)}{\partial t}  &=& -\nabla_1 f_1(s, a, t) - (\gamma(t) + \phi_1 (s, a) + \mu_1(s,a))f_1(s, a, t),\vspace{1.5mm}\\
\dfrac{\partial f_i (s, a, t)}{\partial t}  &=& - \nabla_i f_i(s, a, t) - (\gamma(t) + \phi_i (s, a) + \mu_i(s, a))f_{i}(s, a, t) \nonumber\\
&&{} + \phi_{i-1}(s, a)f_{i-1}(s, a, t), \qquad i = 2, 3, 4,
\end{eqnarray}
with the boundary condition:
\begin{equation}
f_1(s, 0, t) = 2 \displaystyle\int_0^{+\infty}\int_0^{+\infty} p(s, s')\phi_{4}(s', a)f_4(s', a, t)\mathrm{d}a \mathrm{d}s',
\end{equation}
and where
$$\gamma(t) =  \dfrac{1}{N(t)}\frac{\mathrm{d}N(t)}{\mathrm{d}t}$$
denotes the population-level growth rate. 

In the above equations, we always have $s > 0$, $t > 0$, and $a_i >0$. 

In the following analysis, we focus primarily on the evolutionary dynamics of the overall cell size distribution, independent of cell cycle phase and age. To this end, we define the total size distribution as
\begin{equation}
f(s, t) = \sum_{i=1}^4 \int_0^{+\infty} f_i(s, a, t) \mathrm{d} a.
\end{equation}
The corresponding steady-state size distribution is denoted by $f(s)$.

\subsection{Regulation strategies}

In the above equations, the cell growth rates $v_i(s, a)$ and cell size controls $\phi_i(s, a)$ play pivotal roles in cell size regulation.

To model the cell growth rate, we first consider a standard exponential growth model (linear dependence on size):
\begin{equation}
\label{eq:lv}
v_i(s, a) = c s.
\end{equation}
Moreover, to capture fibroblast behavior under environmental constraints, we introduce a population-density-dependent growth rate (PD-GR) that incorporates a deceleration mechanism triggered when the total cell population exceeds a critical threshold. Mathematically, to isolate the specific effect of density-dependent feedback, we normalize the baseline growth capacity such that both models share identical growth kinetics in the low-density regime:
\begin{equation}
\label{eq:hv}
v_i(s, a) = 
\begin{cases}
c s,& N(t) < \bar{N}\\
\dfrac{c}{1+ \left(\frac{1}{\bar{N} K}\sum_{n = 1}^{N(t)} s_n\right)^{m}}s, & N(t) \geq  \bar{N},
\end{cases}
\end{equation}
where $K$ and $m$ are constants, $\bar{N}$ is the population threshold that determines the onset of growth inhibition. Here, $s_n$ represents the size of the n-th cell, and the summation term $\sum_{n=1}^{N(t)} s_n$ denotes the total biomass of the population at time $t$. Biologically, such nonlinear feedback mirrors environmental constraints, such as nutrient limitation or contract inhibition, highlighting how microenvironmental regulation complements intrinsic size-control mechanisms to stabilize population growth.

Experimental studies indicate that the duration of the G1 phase is negatively correlated with birth size and exhibits a minimum temporal threshold for progression \cite{Cadart:2018aa}. In contrast, the duration of the M phase appears to be constant and independent of cell size \cite{Liu:2024aa}. Based on these findings, we model the G1 and M phases as timer-controlled stages with durations $A_1$ and $A_4$, respectively. Specifically, $A_1$ is defined as a function of birth size:
\begin{equation}
A_1 = \max\{A_0, b - k s_b\},
\end{equation}
where $s_b$ denotes the birth size, and $A_0$ specifies the minimum duration of the G1 phase.

For the S and G2 phases, where direct evidence for the distinct control mechanisms is limited, we investigate three canonical regulatory strategies---sizer, adder, and timer---to elucidate their potential roles (Fig. \ref{fig:2}b).

\subsection{Numerical scheme}
Due to the complexity of the PDE system, we implemented an agent-based stochastic simulation algorithm using \texttt{C++} to simulate the dynamics of growth, division, and death. This approach allows for real-time tracking of individual cell states (size, age, and phase) and captures stochastic fluctuations and population heterogeneity. The detailed procedure is outlined in Algorithm \ref{alg:01}.

Specifically, regarding cell division, the mother cell is removed and replaced by two daughter cells, whose cells are determined by independently sampling from the kernel distribution $p(s, s')$ (or equivalently $p(u)$). While biologically division typically conserves instantaneous volume strictly (i.e., $s_1 + s_2 = s_{\mathrm{mother}}$), our independent sampling strategy provides a direct Monte Carlo realization of the boundary condition given in Eq. \eqref{eq:3}, ensuring that the simulated population statistics asymptotically converge to the continuous PDE solution. We verified that this independent sampling strategy yields population size distributions identical to those obtained under a strict mass-conservation splitting strategy.

\begin{algorithm}
\caption{Stochastic simulation for cell growth, division, and death}
\label{alg:01}
\begin{algorithmic}[1]
\State{\textbf{Input:} Cell growth rates (function $v_i(s, a)$), control functions $\phi_i (s, a)$, death rates $\mu_i(s, a)$, daughter cell size distribution $p(u)$, phase transition parameters (e.g., $s_d$ (sizer), $T_d$ (timer), $\Delta s$ (adder)), time step $\Delta t$, and simulation end time $T_{\mathrm{end}}$.}
\State{\textbf{Initialize:} Set $t = 0$, and initial cell number $N = N_0$, with all cells in G1 phase ($n_1 = N_0, n_2 = n_3 = n_4 = 0$). Assign initial sizes randomly and set $a = 0$ for all $i$. }
\For {$t=0$ to $T_{\mathrm{end}}$ in steps of $\Delta t$ }
\State{\textbf{Step 1: Cell fate determination}}
\For {each cell }
\State{Depending on the current phase, compute $\mu_i$, $\phi_i$, and update fate:}
\State{\quad -G1: Evaluate transition to S phase, death, or continued growth.}
\State{\quad -S: Evaluate transition to G2, death, or continued growth.}
\State{\quad -G2: Evaluate transition to M, death, or continued growth.}
\State{\quad -M: Determine whether the cell divides or continues. } 
\EndFor
\State{\textbf{Step 2: System update}}
\For {each cell}
\If {cell dies}
\State{Remove the cell, update $N \gets N - 1$, $n_i \gets n_i - 1$.}
\ElsIf{cell enters the next phase}
\State{Update cell numbers: $n_i \gets n_i - 1$, $n_{i+1} \gets n_i + 1$.}
\ElsIf{cell divides ($i=4$)}
\State{Replace the mother cell with two daughter cells of sizes $s_1 = u_1 s$ and $s_2 = u s$, where ratios $u_1, u_2$ are independently sampled from $p(u)$.}
\State{Set $a = 0$ for daughter cells, update $N \gets N+1$, $n_i \gets n_i - 1$, $n_1 \gets n_1 + 2$.}
\Else
\State{Cell continues in the same phase, update age: $a \gets a + \Delta t$. }
\EndIf
\EndFor
\EndFor
\end{algorithmic}
\vspace{0.25cm}
\noindent\textbf{Note:} Here, the variable $a$ represents the cell cycle age of a cell since its birth. However, the phase-specific rate functions $v_i, \phi_i, \mu_i$ often depend on the residence time within the current phase (phase age). To bridge this, each simulated cell object maintains an auxiliary attribute $a_{\text{entry}}$, recording the value of $a$ at the moment the cell entered its current phase $i$. The effective phase age $\hat{a}_i$ used for evaluating rate functions is computed as $\hat{a}_i = a - a_{\text{entry}}$.
\end{algorithm}

To manage computational cost and memory usage during exponential growth, we employ a downsampling strategy. A global maximum cell number $N_{\max} = 1 \times 10^4$ is predefined. At each update step, if $N(t) > N_{\max}$, we randomly sample $N_{\max}$ cells from the current population to the next iteration. This method preserves population-level heterogeneity while ensuring computational efficiency.

\section{Results}

\subsection{Model foundation}
\subsubsection{Parameter identifiction}
\label{sec:3.1}
\begin{figure}[htbp]
\centering
\includegraphics[width=11cm]{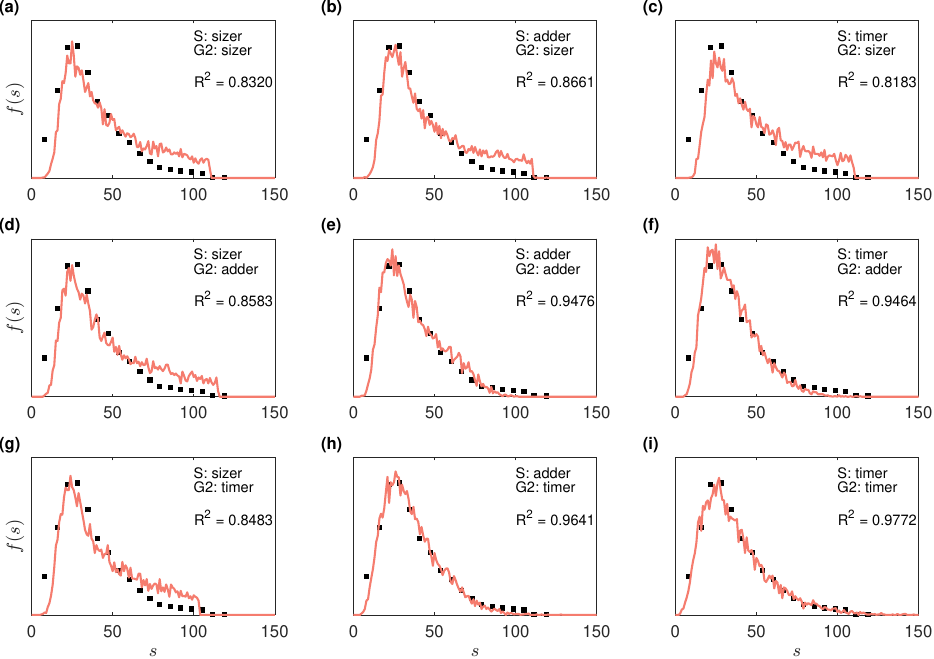}
\caption{Calibration of control mechanisms against experimental fibroblast size distributions. Scatter points represent experimental data from Russel et al. \cite{russell1976cell}. Red solid lines denote simulation results at equilibrium ($t = 400$) using parameters from Table \ref{table1}. For clarity, the specific control strategies adopted for the S and G2 phases, along with the corresponding coefficient of determination ($R^2$), are explicitly labeled in each subplot.}
\label{fig:3}
\end{figure}

To calibrate key model parameters, we performed data fitting using experimental results on fibroblasts reported by Russell et al.  \cite{russell1976cell}. In their study, electrical impedance techniques were employed to characterize fibroblasts derived from normal skin, normal scars, and keloid tissues. Under consistent experimental conditions, different fibroblast types exhibited no significant differences in cell size distributions. Therefore, we used the averaged cell size distribution from these fibroblast lines as a baseline reference for parameter estimation (Fig. \ref{fig:3}). 

To investigate the regulatory mechanisms of the S and G2 phases, we conducted a series of agent-based stochastic simulations using combinations of classical size control strategies---namely, sizer, adder, and timer---for these phases. Each simulation was evaluated by fitting the resulting size distribution to the experimental data, allowing us to identify the optimal control mechanisms and associated parameters.

\begin{table}[h]
\caption{Parameter values used in data fitting (Fig. \ref{fig:3}).}
\label{table1}
\begin{tabular}{@{}ccccccccc@{}}
\toprule
Subplot & $A_0$ & $k$  & $b$ & $A_4$ & S phase (threshold) & G2 phase (threshold) &  $\alpha$ & $\beta$ \\
\midrule
(a) & 2 & 0.1 & 20 & 0.01 & sizer (90) & sizer (110) & 14.4 & 67  \\
(b) & 2 & 0.1 & 25 & 0.01 & adder (35) & sizer (110) & 15.2 & 74 \\ 
(c) & 2 & 0.1 & 10 & 0.01 & timer (10) & sizer (110) & 16 & 72 \\
(d) & 2 & 0.1 & 18 & 0.01 & sizer (90) & adder (20) & 13 & 65  \\ 
(e) & 2 & 0.2 & 20 & 0.01 & adder (20) & adder (15) & 10.1 & 32.5  \\ 
(f) & 2 & 0.1 & 10 & 0.01 & timer (10) & adder (15) & 10.5 & 24 \\ 
(g) & 2 & 0.1 & 25 & 0.01 & sizer (90) & timer (3) & 14 & 65  \\ 
(h) & 2 & 0.1 & 11 & 0.01 & adder (15) & timer (8) & 8.1 & 20  \\
(i) & 2 & 0.1 & 10 & 0.01 & timer (5) & timer (8) & 14.8 & 38 \\ 
\botrule
\end{tabular}
\footnotetext{Key parameters include the negative correlation coefficients $k$ and $b$ for G1 phase duration, the control thresholds for the S and G2 phases, and the shape parameters $\alpha$ and $\beta$ for the daughter size distribution $p(u)$. The minimum G1 phase duration $A_0 = 2$ and the fixed M phase duration $A_4 = 0.01$ are used across all cases. Time is measured in hours; cell size is in arbitrary units.}
\end{table}

In these simulations, cell death was neglected (i.e., $\mu_i = 0$), and the minimum G1 and fixed M phase durations were set to $A_0 = 2$ h and $A_4 = 0.01$ h, respectively. For each combination of control strategies in the S and G2 phases, we tuned the G1 regulation parameters ($k, b$) and the division kernel parameters ($\alpha, \beta$) to minimize the discrepancy between the simulated and experimental size distributions. To isolate the effects of checkpoint mechanisms, additional parameters were fixed: growth rate coefficient $c = 0.1$, inhibition constants $K = 50$ and $m = 1$, and cell number threshold $\bar{N} = 7000$. Here, we adopted the population-density-dependent growth rate (PD-GR) defined in \eqref{eq:hv}, which imposes growth repression at high population densities. A detailed comparison between linear and PD-GR models is presented in subsequent sections.

Figure \ref{fig:3} presents a systematic comparison of model fits against experimental data across nine regulatory strategy combinations. This matrix of results highlights how different mechanisms in the S and G2 phases shape the population size distribution. Broadly, the results fall into two categories. Combinations involving a sizer mechanism in either phase (e.g., Fig. \ref{fig:3}a-d, g) consistently yield poorer fits ($R^2 < 0.9$). As shown in the graphs, the sizer strategy imposes a rigid size threshold, leading to artificial sharp peaks or discontinuities in the distribution that contradict the smooth, unimodal profiles observed empirically.

In contrast, strategiers relying on adder or timer mechanisms (e.g., Fig. \ref{fig:3}e, f, h, i) generate smooth distributions that closely match the experimental data ($R^2 > 0.9$). This improved fit arises because adder and timer controls are inherently more probabilistic and gradual, effectively preserving the natural heterogeneity of the cell population.

Among the evaluated scenarios, two combinations stood out: the ``S(adder)+G2(timer)'' configuration (denoted as (adder, timer), $R^2 = 0.9644$, Fig. \ref{fig:3}h) and the ``S(timer)+G2(timer)'' configuration (denoted as (timer, timer), $R^2 = 0.9660$, Fig. \ref{fig:3}i). Although the pure (timer, timer) model yielded a marginally higher statistical fit, we select the (adder, timer) combination as the baseline configuration for this study. This decision is driven by robustness considerations rather than fitting metrics alone. Theoretically, purely timer-based regulation lacks a size-correction feedback mechanism (i.e., smaller cells do not grow more to catch up), which leads to diverging size variance, particularly under high growth rates or linear growth conditions \cite{jun2015cell,Fan:2025dg}. Our subsequent analysis confirms that the (timer, timer) model fails to maintain homeostasis without the stabilizing constraint of the PD-GR feedback. By contrast, the adder mechanism provides a partial correction for size deviations (adding a constant size), acting as a necessary stabilizer. Therefore, the (adder, timer) configuration represents the most biologically plausible balance between fitting accuracy and long-term dynamical stability.

\subsubsection{Linear growth rate assumption}

In Section \ref{sec:3.1}, we identified model parameters under the nonlinear growth law \eqref{eq:hv}, where a feedback mechanism reduces the cell growth rate once the population size exceeds a threshold. This feedback ensures saturation of growth and prevents unbounded expansion. A fundamental question arises: is such population-level feedback essential for maintaining robust size homeostasis at the single-cell level? To address this, we consider the simpler case of a linear growth law \eqref{eq:lv} and examine how different control mechanisms in the S and G2 phases shape the steady-state size distribution.

\begin{figure}[htbp]
\centering
\includegraphics[width=11cm]{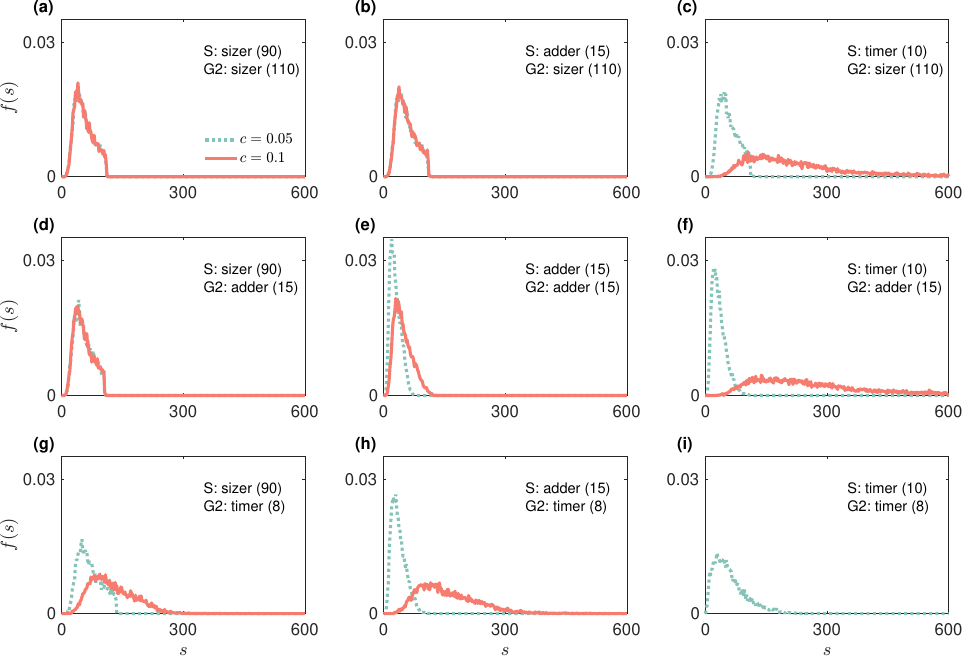}
\caption{Steady-state size distributions under different combinations of S- and G2-phase control mechanisms with linear growth rates. Control strategies along with threshold parameters are shown in each subplot. The cells grow at rate $v_i(s, a) = c s$ with $c = 0.05$ or $0.1$. Unless otherwise specified, other parameters are consistent with row (h) in Table \ref{table1}.}
\label{fig:4}
\end{figure}

Guided by the calibration results in Fig. \ref{fig:3}, we adopted the baseline parameters corresponding to the (adder, timer) configuration (Fig. \ref{fig:3}h and Table \ref{table1}) and performed stochastic simulations of individual cell growth and division. In this setting, cells grow at a linear rate $v_i = c s$, with $c$ denoting the growth constant. While the G1 phase parameters were kept fixed, the control strategies for the S and G2 phases were systematically varied among sizer, adder, and timer mechanisms, using the thresholds specified in Fig. \ref{fig:4}. 

The simulation results (Fig. \ref{fig:4}) reveal four distinct categories regarding steady-state distributions:
\begin{enumerate}
\item \textbf{Growth-rate-insensitive distributions.} For combinations dominated by sizer controls (panels (a), (b), and (d)), the size distribution remains stable and largely unaffected by changes in the growth constant $c$. This stability arises because the sizer mechanism enforces a hard size threshold that overrides variations in growth speed.
\item \textbf{Bypassing of the sizer mechanism.} In cases where the sizer threshold is too high relative to the growth dynamics of previous phases (panels (c) and (g)), cells frequently satisfied downstream regulatory conditions before reaching the sizer threshold. Consequently, the sizer essentially fails to trigger, making the steady-state distribution strongly dependent on $c$.
\item \textbf{Sensitivity of mixed adder/timer mechanisms.} When a timer control operates in at least one phase (e.g., panels (e), (f), and (h)), the distributions became highly sensitive to $c$. Specifically, the mean cell size increases sharply with higher growth rates. This is because, under timer control, the volume added is proportional to the exponential of the product of growth rate and time ($e^{c T_d}$, where $T_d$ represents the timer threshold), leading to cumulative size expansion.
\item \textbf{Instability of dual timer control.} When both the S and G2 phases are regulated by timers (panel (i)), size homeostasis is preserved only at low growth rates. At higher $c$, the distributions broadened substantially, leading to pronounced variability. This confirms theoretical predictions that pure timer control lacks the necessary error-correction mechanism to restrain size divergence \cite{jun2015cell}.
\end{enumerate}

Taken together, these simulations suggest that under a linear growth law, fibroblasts are unlikely to rely on strict sizer-based regulation, a finding consistent with the poor fits observed in Fig. \ref{fig:3}. Instead, adder- and timer-like mechanisms better capture the characteristic balance between homeostasis and heterogeneity. However, a critical limitation emerges: under linear growth assumptions, the steady-state size distribution for these mechanisms remains highly sensitive to the growth constant $c$. This sensitivity implies that without additional feedback (such as the population-density-dependent regulation discussed next), the cell size distribution would lack robustness against fluctuations in growth conditions.

\subsubsection{Nonlinear growth rate assumption}
In the previous section, we examined size control under a linear growth law and found that the steady-state distribution was highly sensitive to the growth constant $c$. To assess whether nonlinear feedback could mitigate this sensitivity, we next investigated cell size distributions under nonlinear growth dynamics. 

\begin{figure}[!ht]
\centering
\includegraphics[width=12cm]{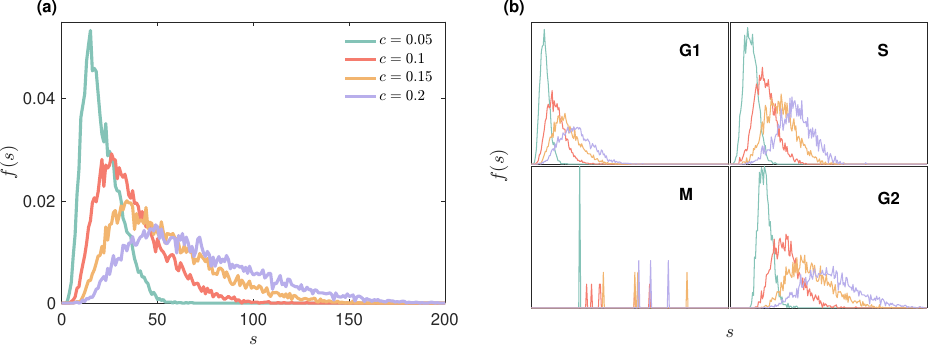}
\caption{Steady-state size distributions under nonlinear growth rate (PD-GR). (a) Steady-state distribution with growth constants $c$ ranging from $0.05$ to $0.2$. (b) Phase-specific distributions.}
\label{fig:5}
\end{figure}

We focused primarily on the (adder, timer) combination, identified earlier as the baseline configuration (Fig. \ref{fig:3}h). Alongside this, we implemented the population-density-dependent growth rate (PD-GR) [Eq. \eqref{eq:hv}] and systematically varied the growth constant $c$. Figure \ref{fig:5}a shows the resulting size distributions for $c$ ranging from $0.05$ to $0.2$. As expected, increasing $c$ shifted the distributions toward larger cell sizes. However, when comparing $c=0.05$ to $0.1$, the PD-GR model exhibited substantially greater robustness than the linear case (Fig. \ref{fig:4}h), resulting in much smaller shifts in the steady-state distributions. Even at higher growth rates ($c = 0.15$ to $c = 0.2$), only minor deviations appeared in the tail of the distribution ($s > 100$). Phase-specific distributions (Fig. \ref{fig:5}b) further revealed that the primary variations occurred in the G1 phase, where the duration depends directly on birth size. Consequently, higher growth rates produced larger birth sizes, and these differences partially propagated into subsequent phases.    

\begin{figure}[htbp]
\centering
\includegraphics[width=12cm]{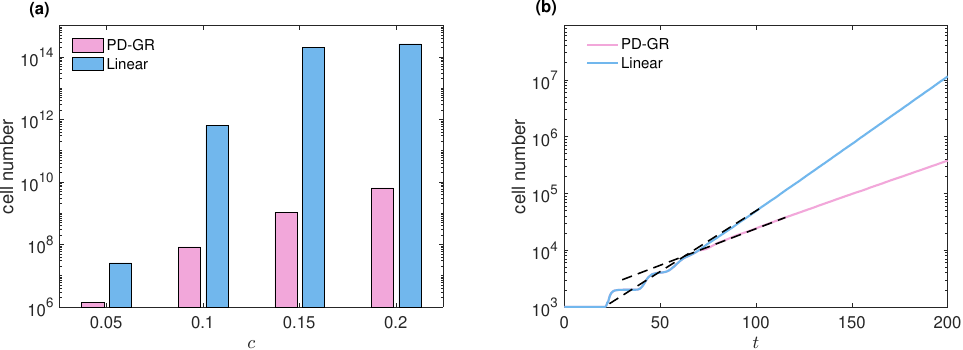}
\caption{Population dynamics. (a) Cell number at $t=400$ under PD-GR versus linear growth, starting from identical initial conditions. (b) Time-course of cell number with $c=0.1$. Dashed lines show the linear approximations. Under PD-GR, feedback slows population expansion once the threshold $\bar{N}$ is reached, mimicking biological constraints such as nutrient limitation or contact inhibition.}
\label{fig:6}
\end{figure}

We further compared linear and PD-GR models at the population level. Starting from identical initial conditions ($N(0) = 1000$), the PD-GR model yielded markedly smaller final cell numbers (Fig. \ref{fig:6}a), reflecting stronger regulations of tissue expansion. Time-course trajectories at $c = 0.1$ (Fig. \ref{fig:6}b) confirmed this pattern: while both growth laws produced rapid initial expansion, the PD-GR trajectory slowed down once $N(t)$ exceeded the threshold $\bar{N}$, consistent with the feedback repression term in \eqref{eq:hv}. This feedback effectively prolonged cycle times---most notably under adder regulation in the S phase---thereby curbing uncontrolled proliferation. To ensure generality, we also evaluated other functional forms of population-density-dependent nonlinear growth (see Appendix \ref{sec:ESNGR}) and found that the choice of specific nonlinear functions did not fundamentally alter the robust characteristics of the cell volume distribution or population size.

Together, these results demonstrate that nonlinear feedback fundamentally reshapes both the steady-state size distribution and the population growth trajectory. Compared with the sensitive and potentially unstable behaviors observed under linear growth, the PD-GR mechanism imposes an extrinsic constraint that reflects microenvironmental factors, such as nutrient limitation and contact inhibition. This highlights the interplay between intracellular mechanisms (adder/timer) and tissue-level feedback.

However, growth rates depend not only on population density but also on cell volume itself, as demonstrated by numerous experiments \cite{BasierNurse:2023cell, Swafferetal:2023RNA,Ginzbergetal:2018cell, DuboisRouzaire-Dubois:2004influence}. Systematic comparisons between population-sensing and volume-sensing regulation remain scarce. To address this, we constructed a cell-size-dependent growth rate (SD-GR), defined as: 
\begin{equation}
\label{eq:v_cellsize}
v_i = \dfrac{cs}{1 + \left({s} /{\hat{s}}\right)^{m_0}},
\end{equation}
where $m_0$ is a scaling exponent and $\hat{s}$ represents the critical size threshold for growth inhibition. We then performed stochastic simulations to compare the phenotypic outcomes of PD-GR and SD-GR mechanisms.

\begin{figure}[htbp]
\centering
\includegraphics[width=12cm]{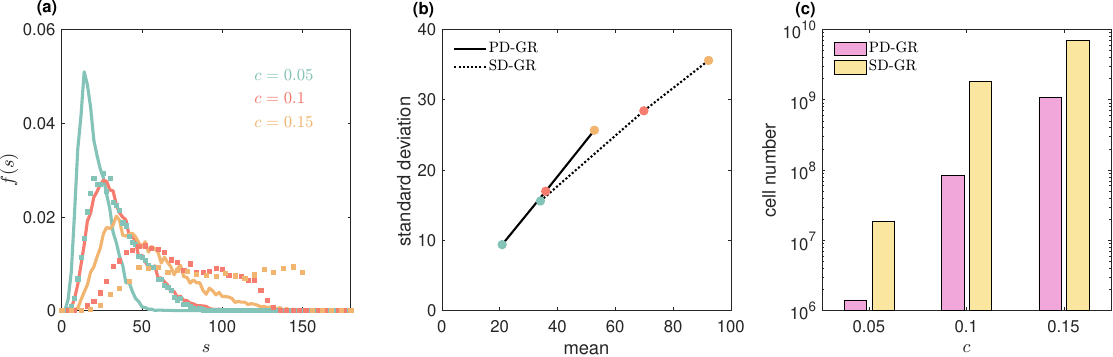}
\caption{Comparative effects of PD-GR and SD-GR on size distributions and cell populations. (a) Steady-state size distributions under PD-GR (solid lines) and SD-GR (dotted lines) for varying $c$. (b) Standard deviation versus mean cell size, where the slope approximates the coefficient of variation (CV). (c) Total cell number at equilibrium ($t = 400$). All simulations used parameter values $m_0 = 6$ and $\hat{s} = 100$ for SD-GR. Note that SD-GR leads to larger cell sizes and higher absolute variability but exhibits a lower relative variability (smaller slope of SD vs mean) compared to PD-GR. }
\label{fig:7}
\end{figure}

Our analysis reveals distinct regulatory signatures for the two mechanisms. Under identical growth parameters $c$, the SD-GR model produces cell size distributions with a larger mean and higher total standard deviation than the PD-GR model (Fig. \ref{fig:7}a). However, when plotting standard deviation against the mean cell size across increasing $c$ values (Fig. \ref{fig:7}b), we observe that the slope of the dependence is noticeably smaller in the SD-GR model compared to PD-GR. This indicates that while SD-GR cells exhibit greater absolute variability, the ratio of standard deviation to mean (coefficient of variation) increases less steeply with growth rate than in PD-GR cells. This suggests that SD-GR establishes a proportional homeostasis mechanism that dampens the growth-rate-induced expansion of relative size variability.

Conversely, at the population level, SD-GR consistently yields a larger final population size than the PD-GR mechanism (Fig. \ref{fig:7}c). Taken together, these results highlight a fundamental trade-off between the two feedback types: PD-GR (extrinsic regulation) prioritizes maintaining a tight population cap, which inherently imposes constraints on achievable cell sizes; while SD-GR (intrinsic regulation) allows for larger absolute growth and population expansion, simultaneously reducing the growth-rate dependence of relative size heterogeneity through size-specific growth inhibition.

\subsection{Parameter sensitivity analysis}
\begin{figure}[htbp]
\centering
\includegraphics[width=11cm]{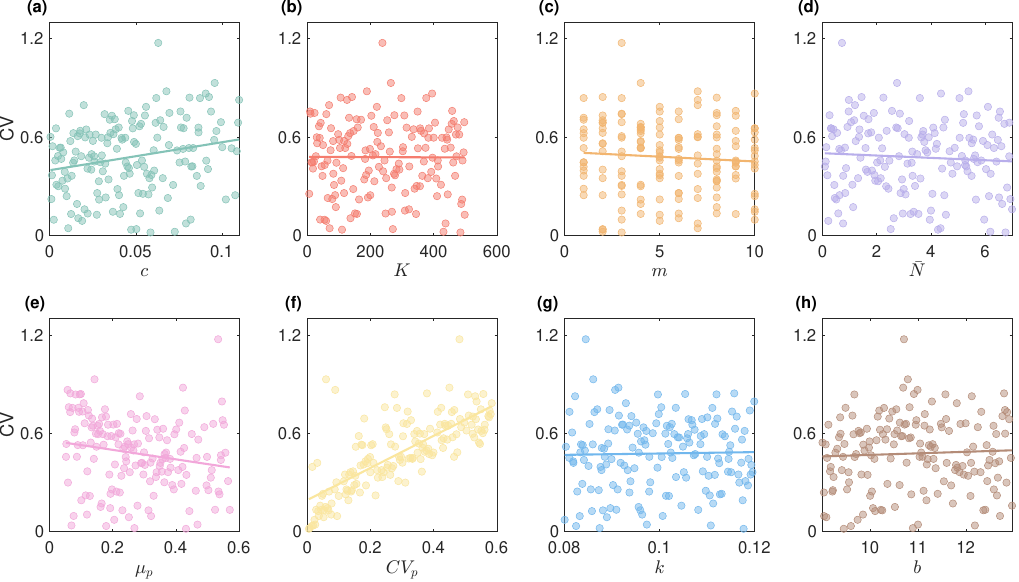}
\caption{Global sensitivity analysis. (a)-(f) show the relationship between steady-state coefficient of variation and sampled values of $c$, $K$, $m$, $\bar{N}$, $\mu_p$, $CV_p$, $k$, and $b$. Scatter points present simulation results; solid lines indicate fitted regression curves. Control mechanisms: adder in the S phase, timer in the G2 phase.}
\label{fig:8}
\end{figure}

To assess the robustness of the steady-state distribution to parameter variability, we conducted a global sensitivity analysis to quantify the relative influence of growth and division parameters on cell-size statistics. Based on the model structure, we selected eight key parameters: the growth rate $c$, nonlinear regulation parameters $K$ and $m$, the threshold cell number $\bar{N}$, the mean $\mu_{p}$ and coefficient of variance $CV_{p}$ of the daughter-size distribution $p(u)$, and the parameters $k$ and $b$ governing G1 duration. Each parameter was sampled across a biologically relevant range using Latin Hypercube Sampling. For each sample set, stochastic simulations were performed, and the steady-state coefficient of variation (CV) of the cell size distribution was recorded. Linear regression was then applied to quantify the CV's sensitivity to parameter variations (Fig. \ref{fig:8}).

The analysis revealed that the growth rate constant $c$, along with the mean $\mu_p$ and coefficient of variation $CV_p$ of the daughter distribution, exerted the strongest influence on the steady-state CV (Fig. \ref{fig:8}). By contrast, parameters $m$ and $\bar{N}$ in the PD-GR term, as well as the G1 phase regulatory parameters $k$ and $b$, had only minor effects. These results align with the recent findings of Proulx-Giraldeau et al. \cite{Proulx-Giraldeau2025:division} in yeast, confirming that population-level size heterogeneity is primarily driven by growth rate and division asymmetry, rather than by phase-specific timing parameters (such as those controlling the G1/S transition).

\begin{figure}[htbp]
\centering
\includegraphics[width=12cm]{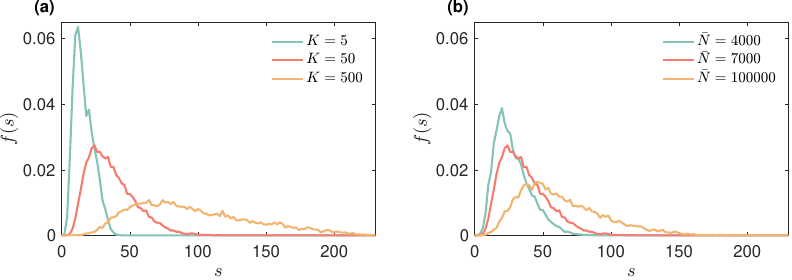}
\caption{Effect of PD-GR feedback parameters on size distribution. (a) Steady-state distributions under different values of $K$. (b) Steady-state distributions under different threshold values $\bar{N}$. Note that larger $K$ or $\bar{N}$ values weaken the effective feedback, leading to broader distributions.}
\label{fig:9}
\end{figure}

Interestingly, the regression analysis suggested that the CV showed little correlation with $K$, which ostensibly contradicts the regulatory role of nonlinear feedback. However, examining the full size distribution reveals substantial changes as $K$ varies from small (strong regulation) to large (weak regulation) values (Fig. \ref{fig:9}a). Larger $K$ values weaken the nonlinear feedback and broaden the tails of the size distribution. This confirms that while $K$ plays a critical regulatory role, its effect is effectively masked when only the CV is considered. Similarly, varying the threshold $\bar{N}$ produced comparable shifts in the steady-state size distribution (Fig. \ref{fig:9}b).

\subsection{Cell regeneration and homeostasis reconstruction}
Under normal physiological conditions, fibroblasts dynamically adjust their cell cycles and respond to microenvironmental cues to maintain tissue homeostasis. In contrast, during tissue injury or inflammation,  fibroblast numbers can be sharply reduced through apoptosis, senescence, or cell cycle arrest, thereby compromising the repair process. To mimic this acute depletion, we performed stochastic simulations in which a large fraction of the cell population was removed and then monitored the subsequent recovery dynamics.

\begin{figure}[htbp]
\centering
\includegraphics[width=12cm]{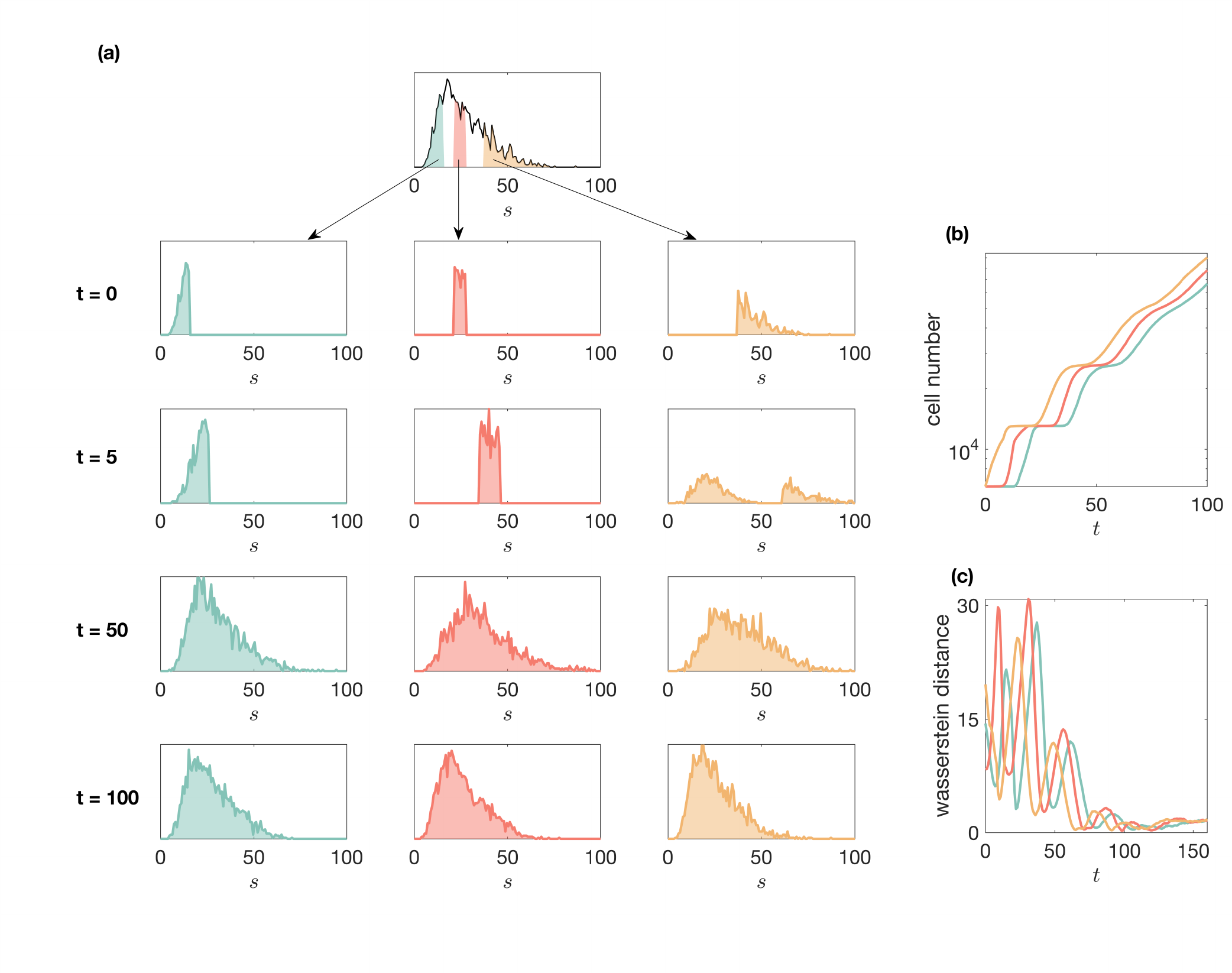}
\caption{Homeostasis reconstruction after cell depletion. (a) Recovery of steady-state size distribution following population depletion under three retention strategies: retaining small cells, intermediate-size cells, or large cells. (b) Population size dynamics after depletion. (c) Recovery of size distribution measured by the Wasserstein distances. }
\label{fig:10}
\end{figure}

In the simulations, once the population size distribution reached a steady state, we retained only $20\%$ of the cells using three distinct retention strategies: (1) small cells, (2) intermediate-sized cells, and (3) large cells. The resulting size distributions differed markedly in the early phase ($t = 5$), reflecting the memory of the perturbation. However, after several cycles ($t = 50$), all scenarios converged toward a common distribution, and by $t = 100$ the original steady-state distribution was fully restored (Fig. \ref{fig:10}a). 

We further examined the recovery dynamics of both population size and distribution. While all three strategies eventually achieved population recovery, retaining large cells resulted in the fastest rebound in total numbers (Fig. \ref{fig:10}b). In our model of actively proliferating fibroblasts, large cells correspond to those in the late stages of the cell cycle (S or G2 phase). Consequently, they possess the highest immediate division potential, allowing for a rapid initial expansion. To quantify the recovery of the size distribution, we computed the Wasserstein distances between the transient distributions and the steady-state reference (Fig. \ref{fig:10}c). The trajectories revealed oscillatory convergence: early-stage fluctuations arose from the discrete nature of synchronized cell division, but these oscillations diminished as the distribution stabilized. Consistent with population-size dynamics, retention of large cells yielded the fastest recovery of distribution likelihood, whereas retention of smaller cells resulted in longer delay before homeostasis was reestablished.

Together, these results highlight the strong regenerative capacity of proliferating fibroblast populations: despite drastic depletion and distinct starting compositions, the system reliably reestablished its steady state. This robustness illustrates the global stability of the steady-state distribution, independent of the initial cell-size composition. Biologically, such resilience provides a theoretical explanation for how healthy fibroblasts can rapidly restore tissue homeostasis after injury, ensuring effective wound healing and functional recovery even when the surviving cell pool is small or compositionally skewed.

\subsection{From in vitro to in vivo: incorporating cell death}
The preceding analysis focused on growth and division dynamics in an expanding population, effectively mimicking \textit{in vitro} exponential growth. However, tissue homeostasis \textit{in vivo} relies on a dynamic equilibrium between proliferation and elimination. To bridge this gap and capture physiological regulation more faithfully, we extended the model to include stochastic cell death and examined its impact on population statistics.

We considered two distinct physiological scenarios: (1) Uniform mortality (cycle-death), where cells experience a constant probability of death throughout the entire cell cycle, representing random environmental insults; and (2) phase-specific vulnerability (SM-death), where death is restricted to the S and M phases, reflecting checkpoints for DNA replication stress or mitotic errors. In both cases, death was modeled with a rate constant $\nu$, incorporated into the stochastic simulation. Simulations were initialized with $N_0 = 5000$ cells, using $\nu = 0.0001$ and $c = 0.1$. Control parameters for cell cycle phases and division kernel parameters ($\alpha$, $\beta$) corresponded to the baseline values in Fig. \ref {fig:3}. 

\begin{figure}[htbp]
\centering
\includegraphics[width=11cm]{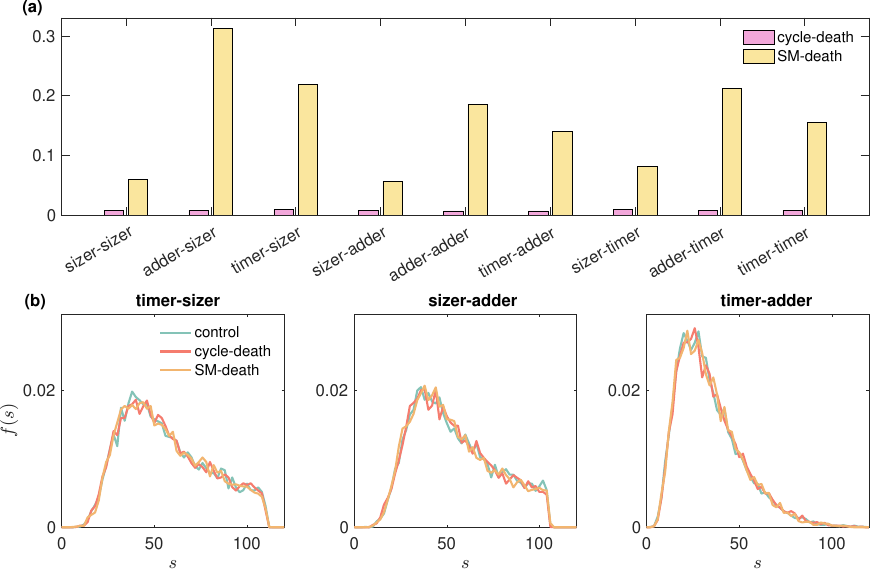}
\caption{Effects of cell death on fibroblast population dynamics. (a) Relative population size under various size-control mechanisms and death assumptions. Values are plotted as the ratio of the cell number with death to the corresponding case without death, under the same control mechanism. (b) Steady-state size distributions under different control mechanisms and death assumptions. ``Control'' indicates the baseline model without cell death, ``cycle-death'' assumes uniform mortarlity across all phases; ``SM-death'' restricts mortality to the S and M phases.}
\label{fig:11}
\end{figure}

Figure \ref{fig:11}a illustrates the relative population sizes under different regulatory combinations. The cycle-death scenario consistently yielded lower population sizes than the SM-death scenario. This is expected, as restricting death to specific phases allows cells to survive the longer G1 intervals. Consequently, under SM-death, population survival becomes highly sensitive to the duration of the vulnerable S-phase relative to the total cycle length.

We then examined the impact of death on the steady-state size distribution (Fig. \ref{fig:11}b). Remarkably, the shape of the distribution remained robust even with the inclusion of cell death. The probability remained robust to the inclusion of cell death. The probability density functions were primarily determined by the underlying size-control strategy (e.g., sizer vs. timer) rather than the death modality. This insensitivity arises because, in our current framework, death is modeled as a size-independent stochastic process. While this simplifies the biological complexity, it reveals an important theoretical distinction: size mechanisms are responsible for maintaining the quality (size homeostasis) of the population, whereas death rates primarily regulate the quantity (population abundance). 

Taken together, these results demonstrate how regeneration efficiency, variability scaling, and death regulation jointly shape fibroblast population dynamics. By incorporating mortality alongside growth and division, the model captures the essential balance of proliferation and loss characterizing living tissues, confirming that size distributions are intrinsic properties of the growth-division machinery, robust to random cellular elimination.

\section{Discussion}
In this study, we developed a segmented mathematical framework to investigate how phase-specific regulatory mechanisms influence population-level cell size dynamics. While the G1 and M phases have been well characterized in the literature, the contributions of the S and G2 phases remain less understood. To bridge this gap, we systematically evaluated classical size-control strategies—sizer, adder, and timer—applied to these intermediate phases and explored their consequences for fibroblast population homeostasis through stochastic simulations.

A key innovation of our work is the comparative analysis of growth feedback mechanisms. Beyond the classical linear growth model, we introduced two physiologically relevant regulation modes: population-density-dependent growth (PD-GR) and cell-size-dependent growth (SD-GR). Our comparison reveals a fundamental trade-off between population capping and size uniformity. The PD-GR mechanism, which represents extrinsic constraints such as contact inhibition, tightly bounds total cell number but permits greater relative size variability. In contrast, the SD-GR mechanism, functioning as an intrinsic soft sizer, significantly dampens the growth-rate dependence of size heterogeneity (lower CV slope) but allows for a more flexible population expansion. This suggests that mammalian fibroblasts may utilize a sophisticated combination of extrinsic and intrinsic feedback to balance tissue space constraints with individual size homeostasis. It is worth noting that while different nonlinear growth patterns imply distinct regulatory logic, they did not result in divergent core properties of the steady-state distributions (Appendix \ref{sec:ESNGR}).

Within the nonlinear framework, sensitivity analysis highlighted that steady-state outcomes are governed by a specific subset of parameters. The growth constant $c$ and the division kernel $p(u)$ emerged as the dominant factors. Specifically, as the mean of $p(u)$ increases, the coverage range of the distribution broadens while the coefficient of variation decreases. Conversely, reducing the variance of $p(u)$ significantly enhances cell size uniformity (Appendix \ref{sec:Diff_p}). We also observed that the regulation stiffness parameter $K$ modulates the breadth of distributions: weaker feedback (larger $K$) enables greater variability even when intrinsic division noise remains constant. These findings suggest that precise control over the division kernel is more critical for size homogeneity than the strictness of growth inhibition feedback.

Perturbation experiments simulating acute fibroblast depletion demonstrated the system's robust regenerative capacity. Our model predicts that retaining large cells accelerates population recovery, a consequence of the active proliferation assumption, in which large size correlates with proximity to mitosis (G2/M phase). We note that this finding contrasts with recent experimental work by Lengefeld et al. \citep{Lengefeld:2021aa}, who reported that enlarged hematopoietic stem cells exhibit reduced regenerative potential. This divergence highlights the context-dependence of cell size physiology: Legngefeld et al. focused on stem cell enlargement associated with aging and functional decline (senescence), whereas our model captures the dynamics of healthy, actively cycling fibroblasts where size accumulation is coupled to division competence. Therefore, our results are most applicable to hyper-proliferative phases, such as early wound healing, rather than long-term stem cell exhaustion. Future model extensions incorporating a senescence-size feedback loop would allow for a unifying theory linking hypertrophy-induced aging with proliferation-driven regeneration.

Extending the model to incorporate cell death revealed that while population numbers are highly sensitive to mortality rates, the shape of size distributions remains remarkably robust. Uniform death (cycle-death) uniformly suppresses population size, whereas phase-specific death (SM-death) amplifies the selection pressure of cell-cycle checkpoints. This insensitivity of size distribution to random death implies that size homeostasis is structurally encoded in the growth-division rules, distinct from the survival-death balance. However, coupling these processes is essential to capture realistic \textit{in vivo} dynamics, in which tissue mass is maintained by a delicate balance between proliferation and apoptosis.

Several simplifying assumptions warrant discussion. First, specific molecular effects (e.g., Rb, Cyclins) were abstracted into phenomenological control functions. Second, the death rate was modeled as a size-independent constant. In reality, as suggested by the study of Lengefeld et al., extremely large cells may be preferentially eliminated via stress response pathways. Implementing a size-dependent death rate could theoretically trim the heavy tail of the distributions observed in our simulations. Third, the growth saturation was instantaneous, simplifying the likely gradual mechanical and chemical signaling in tissues.

In conclusion, this study establishes a comprehensive framework for dissecting how biological trade-offs---between intrinsic vs. extrinsic growth feedback, and between proliferation vs. size control---shape tissue homeostasis. Through stochastic simulation, we reveal the stabilizing power of nonlinear growth feedbacks and the complementary roles of adder- and timer-like mechanisms in the S/G2 phases. 

Looking forward, this framework provides a theoretical basis for interpreting high-resolution data from live-cell imaging or single-cell sequencing. By aligning simulation outputs with experimental distributions, the model can help disentangle cell-intrinsic controls from microenvironmental influences in fibroblast or cancer. Moreover, merging this size-based framework with aging-based models could provide deeper insights into how tissues balance rapid regeneration with the long-term risk of senescence and functional decay.

\bmhead{Acknowledgements}
This work was funded by the National Natural Science Foundation of China (NSFC 12331018).

\bmhead{Data availability statement}
All data used in this study are derived from published sources and model simulations. The experimental datasets are available from the original references cited in the paper. All simulation results can be fully reproduced using the numerical schemes described herein.

\begin{appendices}

\section{Derivation of the governing equations (Eq. \eqref{eq:1}--\eqref{eq:2})}
\label{sec:Derivation_eq}
The cell cycle is partitioned into four distinct stages: G1, S, G2, and M. Let $n_i(s, a, t)$ ($i = 1,2,3,4$) denote the cell number density in the $i$-th stage at time $t$, characterized by cell size $s$ and cell cycle age $a$. The age variable $a$ represents the time elapsed since the cell's birth (entry into G1), and is effectively inherited as the cell transitions between phases. Thus, $a$ is a continuous variable that increases monotonically throughout the entire cycle.

To derive the transport equations, we consider a control volume $D = [s_{1},s_{2}] \times [a_{1}, a_{2}]$ in the state space $(s, a)$. By the principle of number conservation, the change in cell number within $D$ during a time interval $\Delta t$ is determined by the net flux across the boundaries and the source/sink terms:
\begin{equation}
\label{eq:A1}
\begin{aligned}
\Delta n_i(s, a, t) &= \iint\limits_{D} n_i(s, a, t+\Delta t) \mathrm{d} \sigma - \iint\limits_{D} n_i(s, a, t) \mathrm{d}\sigma\\
&= \Delta t \iint\limits_{D} \dfrac{\partial n_i(s, a, t)}{\partial t}\mathrm{d} \sigma + o(\Delta t),
\end{aligned}
\end{equation}
where $\mathrm{d} \sigma = \mathrm{d} s \mathrm{d} a$. The dynamics are driven by: (1) convective flux due to growth in size and aging in time; (2) transition out of the current stage determined by the rate function $\phi_i(s, a)$; (3) influx from the previous stage; and (4) removal due to cell death $\mu_i(s, a)$. 

Focusing on the G1 phase ($i = 1$) where inflow occurs strictly at $a = 0$, the integral conservation law for an arbitrary internal region $D$ is:
\begin{equation}
\label{eq:A2}
\begin{aligned}
&\iint\limits_{D} (n_1(s, a, t+\Delta t)  -  n_1(s, a, t)) \mathrm{d}\sigma\\
	=& \underbrace{\int_{s_{1}}^{s_{2}} \int_{a_{1} - \Delta a}^{a_{1}}   n_1(s, a, t) \mathrm{d} s \mathrm{d} a}_{\text{Aging influx}} - \underbrace{\int_{s_{1}}^{s_{2}} \int_{a_{2} - \Delta a}^{a_{2}}   n_1(s, a, t) \mathrm{d} s \mathrm{d} a}_{\text{Aging efflux}} \\
	&{} + \underbrace{\int_{s_{1}-\Delta s}^{s_{1}} \int_{a_{1}}^{a_{2}}   n_1(s, a, t) \mathrm{d} s \mathrm{d} a}_{\text{Growth influx}} - \underbrace{\int_{s_{2} - \Delta s}^{s_{2}} \int_{a_{1}}^{a_{2}}   {n}_1(s, a, t) \mathrm{d} s \mathrm{d} a}_{\text{Growth efflux}} \\
	&{} - \underbrace{\Delta t \iint\limits_{D} (  {\phi}_1(s,  a) +   {\mu}_1(s,  a))   {n}_1(s,  a,  t) \mathrm{d} \sigma}_{\text{Transition \& cell death}}
\end{aligned}
\end{equation}
Applying the mean value theorem and letting $\Delta t \to 0$ (we note $\Delta a = \Delta t$), we define the instantaneous growth rate $v_1(s, a) = \frac{\mathrm{d} s}{\mathrm{d} t}$ for cells at G1 phase. The differences simplify to partial derivatives:
\begin{eqnarray}
\label{eq:A3}
&&\int_{s_{1}}^{s_{2}} \int_{a_{1} - \Delta a}^{a_{1}}   n_1(s, a, t) \mathrm{d} s \mathrm{d} a - \int_{s_{1}}^{s_{2}} \int_{a_{2} - \Delta a}^{a_{2}}   n_1(s, a, t) \mathrm{d} s \mathrm{d} a\nonumber \\
& = &-\Delta t \iint\limits_{D} \dfrac{\partial n_1(s, a, t)}{\partial a}\mathrm{d} \sigma + o(\Delta t)\\
\label{eq:A4}
&&\int_{s_{1}-\Delta s}^{s_{1}} \int_{a_{1}}^{a_{2}}   n_1(s, a, t) \mathrm{d} s \mathrm{d} a - \int_{s_{2} - \Delta s}^{s_{2}} \int_{a_{1}}^{a_{2}}   {n}_1(s, a, t) \mathrm{d} s \mathrm{d} a \nonumber\\
& = &-\Delta t\iint\limits_{D}\dfrac{\partial (v_1(s, a) n_1(s, a, t))}{\partial s} \mathrm{d} \sigma + o(\Delta t).
\end{eqnarray}
Hence, Eqs. \eqref{eq:A1}--\eqref{eq:A4} together give:
$$
\begin{aligned}
\iint\limits_{D} \dfrac{\partial n_1(s, a, t)}{\partial t}\mathrm{d} \sigma = &-\iint\limits_{D} \dfrac{\partial n_1(s, a, t)}{\partial a}\mathrm{d} \sigma - \iint\limits_{D}\dfrac{\partial (v_1(s, a) n_1(s, a, t))}{\partial s} \mathrm{d} \sigma\\
&{} - \iint\limits_{D} (  {\phi}_1(s,  a) +   {\mu}_1(s,  a))   {n}_1(s,  a,  t) \mathrm{d} \sigma.
\end{aligned}
$$
Since $D$ is arbitrary, the integrand must vanish, yielding the McKendrick-von Foerster type equation for the G1 phase:
\begin{equation}
\dfrac{\partial n_1}{\partial t} + \dfrac{\partial n_1}{\partial a} + \dfrac{\partial \left(v_1(s, a) n_1\right)}{\partial s} = - (\phi_1(s, a) + \mu_1(s, a)) n_1.
\end{equation}

For subsequent stages ($i = 2, 3, 4$), since the age $a$ is continuous, cells entering phase $i$ from phase $i-1$ do so at the same age $a$ they left phase $i-1$. Thus, the transition appears as a distributed source term within the domain, rather than a boundary condition. The conservation law, including an influex term from the previous phase, becomes:
\begin{equation}
\label{eq:dev_others}
\begin{aligned}
\iint\limits_{D} \dfrac{\partial n_i}{\partial t} \mathrm{d}\sigma =& -\iint\limits_{D} \dfrac{\partial n_i}{\partial a}\mathrm{d}\sigma - \iint\limits_{D} \dfrac{\partial (v_i(s, a) n_i)}{\partial s} \mathrm{d}\sigma \\
&{} - \iint\limits_{D}  (\mu_i(s, a) + \phi_i(s, a)) n_i \mathrm{d} \sigma + \iint\limits_{D} \phi_{i-1}(s, a) n_{i-1}(s, a, t) \mathrm{d} \sigma.
\end{aligned}
\end{equation} 
Here, the term $\phi_{i-1}n_{i-1}$ represents cells transitioning into phase $i$ from $i-1$. Removing the integrals yields:
\begin{equation}
\dfrac{\partial n_i }{\partial t}  + \dfrac{\partial n_i}{\partial a} + \dfrac{\partial (v_i(s, a) n_i)}{\partial s} =   - (\mu_i(s, a) + \phi_i(s, a))n_{i}  + \phi_{i-1}(s, a) n_{i-1}, \quad i = 2, 3, 4.
\end{equation}

\section{Derivation of the boundary condition}
\label{sec:Derivation_BC}
The boundary condition applies only to the G1 phase at age $a = 0$, representing the birth of new cells from the division of M-phase mother cells. The total flux of dividing mother cells (leaving phase 4 via $\phi_4$) involves integrating over all possible ages $a$ and sizes $s'$:
$$
J_{\text{division}}(t) = \int_0^{+\infty} \int_{0}^{+\infty} \phi_4(s', a) n_4(s', a, t) \mathrm{d} s' \mathrm{d} a.
$$

Considering the division kernel $p(s, s')$, and that one mother produces 2 daughter cells upon division, the population density of newborn daughter cells with size $s$ at age $a = 0$ is:
\begin{equation}
n_1(s, 0, t) = 2\int_0^{+\infty} \int_{0}^{+\infty} p(s, s') \phi_4(s', a) n_4(s', a, t) \mathrm{d} s' \mathrm{d} a.
\end{equation}
For phase $i=2, 3, 4$, there are no boundary conditions at $a = 0$ since cells cannot be born directly into these stages; they enter solely via transition from the preceding phase at $a > 0$, which is captured by the source terms in the governing equations.

\section{Extended simulation of nonlinear growth rates}
\label{sec:ESNGR}
While the preceding discussion highlighted the significance of the threshold-dependent population-density growth model (referred to as PD-GR in the main text), density regulation in biological systems often exhibits diverse nonlinear characteristics. Distinct functional forms of feedback may lead to varied population dynamics. To investigate the robustness of our findings and the effects of alternative regulation architectures, we introduced two additional nonlinear growth rate models---Global Biomass Feedback (GB-GR) and biphasic adaptive feedback (BAF-GR)---and systematically compared their regulatory effects against the baseline model.

The first alternative model, termed Global Biomass Feedback (GB-GR), assumes that cells are continuously constrained by population-level negative feedback throughout the entire growth process, driven by the total biomass rather than cell number. It is expressed as:
\begin{equation}
\label{Aeq:GB-GR}
v_i = \dfrac{c s}{1+ \left( \frac{1}{\bar{N} K}\sum_{n=1}^{N(t)}s_n\right)^m},
\end{equation}
where the definitions of parameters $c$, $\bar{N}$, $K$, and $m$ are identical to those in Eq. \eqref{eq:hv}. Unlike the threshold-dependent model, GB-GR imposes a global metabolic constraint without a distinct transition phase.

The second model is the Biphasic Adaptive Feedback (BAF-GR). This formulation introduces a dual-regime mechanism: a weak feedback phase at low densities mimicking initial resource sensing, followed by strong inhibition once the capacity is reached. It is defined as:
\begin{equation}
\label{Aeq:BAF-GR}
v_i= 
\begin{cases}
\dfrac{c s}{1 + \epsilon \left( \frac{N(t)}{\bar{N}}\right)^l },\quad N(t) < \bar{N},\vspace{1.5mm}\\
\dfrac{c s}{1+ \left( \frac{1}{\bar{N} K}\sum_{n=1}^{N(t)} s_n\right)^m}, \quad N(t) \geq \bar{N},
\end{cases}
\end{equation}
where $\epsilon$ represents the weak feedback coefficient, and $l$ denotes the Hill exponent governing the nonlinearity of the weak feedback phase. Notably, when the population $N(t)$ exceeds the threshold $\bar{N}$, the growth rate converges to an independent nonlinear suppression similar to the strong feedback phase of the PD-GR.

\begin{figure}[htbp]
\centering
\includegraphics[width=12cm]{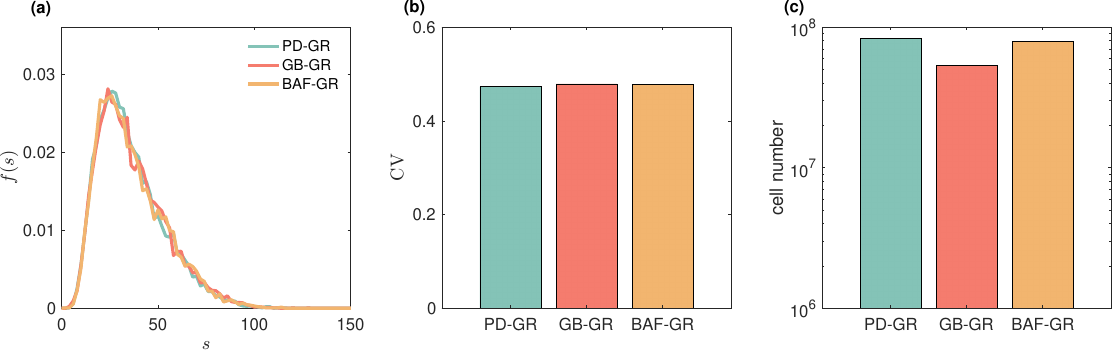}
\caption{Properties of cell populations governed by varying nonlinear growth feedback mechanisms. (a)--(c) Comparison of steady-state cell size distribution, coefficient of variation (CV), and total cell number among three growth models at $t = 400$. The Global Biomass Feedback is defined by Eq. \eqref{Aeq:GB-GR}, while the biophasic adaptive feedback (BAF-GR) is described by Eq. \eqref{Aeq:BAF-GR}.}
\label{fig:12}
\end{figure}

Simulation results indicate that the specific functional form of nonlinear feedback has a negligible impact on the intrinsic phenotypic characteristics of the population, such as size distribution and coefficient of variation (Fig. \ref{fig:12}a--b). The primary distinction lies in the regulatory intensity: different feedback mechanisms lead to variations in proliferation efficiency, which are most pronounced in the final steady-state population size (Fig. \ref{fig:12}c).

\section{Impact of division kernel parameters on population heterogeneity}
\label{sec:Diff_p}
To systematically investigate how the stochasticity of cell division affects population-level dynamics, we analyzed the sensitivity of the system to the shape parameters ($\alpha, \beta$) of the beta distribution $p(u)$. We constructed four distinct parameter sets to decouple the effects of division asymmetry (mean determination) from division noise (variance):
\begin{itemize}
\item Set 1 (Baseline Asymmetry): Parameters $\alpha = 8.1$ and $\beta = 20$ (derived from Table \ref{table1}, row h), representing a biologically typical asymmetric division scenario. 
\item Set 2 \& 3 (Variance-Constrained mean shift): We set $(\alpha, \beta) = (17.11, 17.11)$ and $(\alpha, \beta) =(20, 8.1)$, respectively. These sets maintain the same variance as Set 1 but shift the mean division ratio $\mu_p$, allowing us to test the effect of symmetry versus asymmetry without altering the intrinsic noise level.
\item Set 4 (High Precision): We set $\alpha = \beta = 96$. This configuration maintains a symmetric mean ($\mu_p = 0.5$) but significantly reduces the variance, mimicking highly precise (low-noise) cell division.
\end{itemize}

\begin{figure}[htbp]
\centering
\includegraphics[width=12cm]{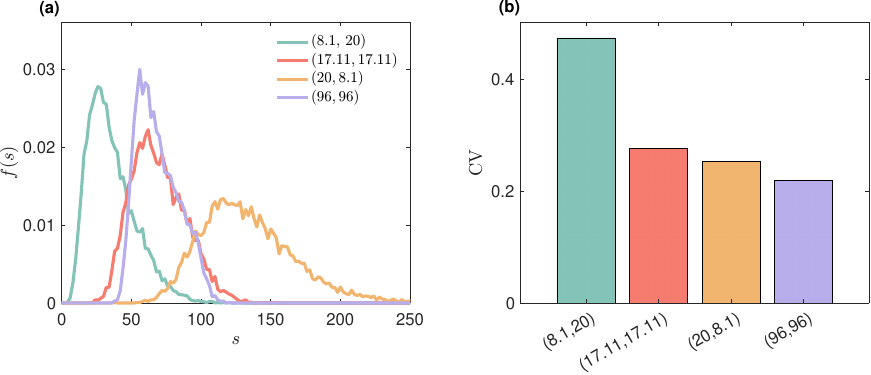}
\caption{The influence of division kernel's shape parameters $(\alpha, \beta)$ on cell population statistics. (a) Steady-state cell size distributions at $t = 400$ across four parameter configurations: $(8.1, 20)$, $(17.11, 17.11)$, $(20, 8.1)$, and $(96, 96)$. The first three sets preserve a constant variance in the division ratio $u$, whereas the fourth set ($\alpha = \beta = 96$) represents a low-variance, symmetric division regime. (b) Comparison of the coefficient of variation (CV) of cell size for the corresponding parameter sets.}
\label{fig:13}
\end{figure}
The simulation results reveal two critical insights into how division properties shape population heterogeneity. First, deviations from symmetric division (i.e., shifting $\mu_p$ away from $0.5$) tend to broaden the cell size distribution range, even when the division variance remains constant. In contrast, under symmetric division ($\mu_p = 0.5$,  achieved when $\alpha = \beta$), the peak position of the size distribution remains stable regardless of the noise level (Fig. \ref{fig:13}a). Second, the precision of division plays a dominant role in regulating heterogeneity: increasing the magnitude of $\alpha$ and $\beta$ (Set 4) drastically suppresses the variance of the division ratio. This reduction in division noise directly translates to a more concentrated cell size distribution and a lower population CV (Fig. \ref{fig:13}b), highlighting that precise daughter size allocation is essential for maintaining size uniformity.

\end{appendices}

\bibliography{references}

\end{document}